\documentclass[onecolumn]{revtex4}%
\usepackage{amssymb}
\usepackage{amsmath}
\usepackage{epsfig}
\usepackage{amsfonts}
\usepackage{graphicx}%
\begin{document}
\title{Radiative $\phi$ decays with derivative interactions}
\author{Francesco Giacosa and Giuseppe Pagliara}
\affiliation{Institut f\"{u}r Theoretische Physik, Johann Wolfgang Goethe University,
Max-von-Laue-Str.~1, D-60438 Frankfurt, Germany }

\begin{abstract}
We study the line shapes of radiative $\phi$-decays involving virtual
$f_{0}(980)$ and $a_{0}(980)$ mesons which decay, via derivative couplings, to
$\pi^{0}\pi^{0}$ and $\pi^{0}\eta$ respectively. After developing the
formalism for derivative interactions at one-loop level, we show that they can
reproduce the measured peaked line shapes of $\phi$-decays without including
kaon loops.

\end{abstract}
\maketitle









\section{Introduction}

The radiative decays of the $\phi$ meson are a valuable tool to study the
nature of light scalar states below 1 GeV \cite{amslerrev,achasovorig,close}.
It is still debated if the latter are quarkonia \cite{scadron}, tetraquark
\cite{tqgen,tq}, molecular states \cite{mesonicmol1} or a mixing of these
configurations \cite{fariborz,tqmix}. A dominant quarkonium assignment is
problematic \cite{noqq}, thus leading to identify the $\overline{q}q$-states
with resonances above 1 GeV where mixing with the scalar glueball takes place
\cite{amslerclose,refs}.

In this article we do not employ a particular interpretation for the
resonances $f_{0}\equiv f_{0}(980)$ and $a_{0}\equiv a_{0}(980)$.
Independently on their nature, spontaneous breaking of chiral symmetry implies
that derivative couplings to the pseudoscalar mesons $\pi,$ $\eta$ and $K$
arise. In fact, the latter states are the emerging Goldstone bosons, which can
be rewritten as angular excitations. Then, Lagrangian interactions of
derivative type with schematic form $\mathcal{L}_{int}\sim S\left(
\partial_{\mu}\varphi_{1}\right)  \left(  \partial^{\mu}\varphi_{2}\right)  $
with $S=f_{0},a_{0}$ and $\varphi_{1,2}=\pi,\eta,K$, are obtained
\cite{ecker}. Our aim is to study in detail the effects of such derivative
interactions on the line shapes of the reactions $\phi\rightarrow\gamma\pi
^{0}\pi^{0}$ and $\phi\rightarrow\gamma\pi^{0}\eta$, occurring via virtual
$f_{0}$ and $a_{0}$ mesons respectively.

In order to discuss properly the issue we concentrate on the reaction
$\phi\rightarrow\gamma\pi^{0}\pi^{0}$ via a virtual $f_{0}$ state. We first
write down explicitly the general interaction Lagrangian of the $f_{0}$ meson
with pions and kaons as emerging upon chiral symmetry breaking within the
nonlinear realization of chiral symmetry as obtained in Ref. \cite{ecker}:%
\begin{equation}
\mathcal{L}_{int,f_{0}}=c_{f_{0}\pi\pi}f_{0}\left(  \partial_{\mu
}\overrightarrow{\pi}\right)  ^{2}+d_{f_{0}\pi\pi}M_{\pi}^{2}f_{0}%
\overrightarrow{\pi}^{2}+c_{f_{0}KK}f_{0}\left(  \partial_{\mu}K^{+}\right)
\left(  \partial^{\mu}K^{-}\right)  +d_{f_{0}KK}M_{K}^{2}f_{0}(K^{+}K^{-})+...
\label{f01}%
\end{equation}
where dots refer to the analogous terms with the neutral kaon states. In the
chiral limit the pion and the kaon masses $M_{\pi}$ and $M_{K}$ vanish leaving
only interactions with derivatives parametrized by the coupling constants
$c_{f_{0}\pi\pi}$ and $c_{f_{0}KK}$. The constants $d_{f_{0}\pi\pi}$ and
$d_{f_{0}KK}$ parametrize the interaction without derivatives. Now, the
radiative decay $\phi\rightarrow\gamma\pi^{0}\pi^{0}$ via $f_{0}$ can occur
essentially in two ways:

(a) An interaction Lagrangian for the decay $\phi\rightarrow\gamma f_{0}$ is
introduced as $\mathcal{L}_{\phi\gamma f_{0}}=c_{\phi\gamma f_{0}}f_{0}%
F^{\mu\nu}V_{\mu\nu},$where $F^{\mu\nu}=\partial^{\mu}A^{\nu}-\partial^{\nu
}A^{\mu}$ is the electromagnetic field strength and $V_{\mu\nu}=\partial_{\mu
}\phi_{\nu}-\partial_{\nu}\phi_{\mu}$ is the field strength related to the
vector field $\phi_{\mu}.$ The interaction Lagrangian $\mathcal{L}_{\phi\gamma
f_{0}},$ parameterized by $c_{\phi\gamma f_{0}},$ corresponds to a point-like
coupling which effectively takes into account loops of quarks (whose precise
form depends on the microscopic interpretation of the $f_{0}$ meson). After
the transition $\phi\rightarrow\gamma f_{0}$ the scalar meson $f_{0}$ decays
into pions (as depicted in Fig. 1.c) via derivative couplings (mechanism a.1,
whose amplitude is proportional to $c_{f_{0}\pi\pi})$ or via non-derivative
couplings (mechanism a.2, proportional to $d_{f_{0}\pi\pi}$). A first study of
derivative interactions (thus setting $d_{f_{0}\pi\pi}=d_{f_{0}KK}=0$) has
been performed in Ref. \cite{black}, where it has been shown by using
Breit-Wigner propagators for the scalars that derivative couplings describe
the data better than non-derivative ones. A detailed study involving
non-derivative interactions only (setting $c_{f_{0}\pi\pi}=c_{f_{0}KK}=0)$ has
been performed in Ref. \cite{isidori}.

(b) The vector meson $\phi$ couples strongly to kaons. Then, via a kaon-loop a
photon is generated at one vertex and a $f_{0}$ meson at the other vertex. The
latter coupling can again occur in two ways: via derivative coupling
(mechanism b.1, proportional to $c_{f_{0}KK},$ see Ref. \cite{black}) or via
non derivative coupling (mechanism b.2, proportional to $d_{f_{0}KK},$ see
Ref. \cite{achasovorig,achasov,gubin}).

Note that all these considerations apply also for the $a_{0}$ meson as a
virtual state of the reaction $\phi\rightarrow\gamma\pi^{0}\eta$: the
derivative (non-derivative) interactions with $\pi\eta$ and $\overline{K}K$
are parameterized by $c_{a_{0}\pi\eta}$ ($d_{a_{0}\pi\eta}$) and $c_{a_{0}KK}$
($d_{a_{0}KK}$) in a Lagrangian which is analogous to Eq. (\ref{f01}).
Clearly, the possible decay mechanisms are also separated into a.1 and a.2
(direct, non-structure coupling) and via kaon loops (b.1 and b.2).

It is evident that the description of the radiative decay $\phi\rightarrow
\gamma\pi^{0}\pi^{0}$ (and $\phi\rightarrow\gamma\pi^{0}\eta)$ is difficult
because four decay mechanisms (a.1, a.2 and b.1, b.2) can potentially
contribute and it is not clear $a$ $priori$ which one is dominant. As a
consequence, considering that the inclusion of all contributions at the same
time has not yet been performed, the extraction of the parameters of Eq.
(\ref{f01}) from experiments depends also on theoretical assumptions.

Mechanism (b) is a mesonic 1-loop contribution to the decay mechanism, which
is regarded as dominant by many authors in view of the large coupling to kaons
of the vector meson $\phi$ meson and of the scalars $a_{0}(980)$ and
$f_{0}(980).$ However, notice that the mechanism (a) is dominant according to
large-N$_{c}$ counting rule both in the tetraquark and quarkonium assignments
for the $f_{0}(980)$ meson and in the tetraquark assignment for the
$a_{0}(980)$ \cite{thooft} (see also the note to this Reference). While it is
not yet clear to which extent large-N$_{c}$ is reliable in this context -see
also Ref. \cite{close} for a discussion of this point-, we consider this fact
as a motivation to study in detail the effects of the mechanism (a) on
radiative decays and thus to test an alternative scenario for the description
of radiative $\phi$ decays. Moreover, in this work we restrict to the chiral
limit dominant mechanism a.1 where only derivatives are involved
(corresponding to setting $d_{f_{0}\pi\pi}=d_{f_{0}KK}=d_{a_{0}\pi\eta
}=d_{a_{0}KK}=0$). We thus intend to continue the analysis initiated in Ref.
\cite{black} about derivative interactions by studying the decay mechanism a.1
in both the $f_{0}$ and the $a_{0}$ channels in relation to radiative $\phi$
decays. We aim to do it by properly taking into account loops and finite-width
effects using the formalism developed in Ref. \cite{lupo} extended to the case
of derivative couplings. That is, both real and imaginary parts of self-energy
contributions (which show a rather different behavior than their
non-derivative counterparts, see Fig. 2) are taken into account. We regularize
the model by using an effective cutoff of the order of $1$ GeV which is
introduced by using a nonlocal extension of Eq. (\ref{f01}). We conclude this
discussion by noting that the neglect of the non-derivative coupling is surely
justified in virtue of the small pion mass in the $\pi\pi$ channel of $f_{0}$
but is less justified in $\pi\eta$ channel of $a_{0}$ and in the kaon-kaon
channel of both resonances because of the larger masses of the $K$ and $\eta$
mesons. The next step shall be the inclusions of mechanism a.2. but at the
present stage a fit with non-zero $c$'s and $d$'s at the same time would not
be constrained enough.

Quite remarkably, our analysis shows that derivative interactions alone work
well in the description of both $\phi\rightarrow\gamma\pi^{0}\pi^{0}$ and
$\phi\rightarrow\gamma\pi^{0}\eta$ line shapes as experimentally measured in
the SND and KLOE collaborations in Refs. \cite{sndf0,kloef0} for the $f_{0}$
meson and in Refs. \cite{snda0,kloea0} for the $a_{0}$ meson. The peaked line
shapes can be reproduced in virtue of the derivative coupling which enhance
the theoretical curves close to threshold.

We also compare our results with Ref. \cite{otherf0}, where the $f_{0}$ meson
has been studied studied in $j/\psi$ decays at BES, and with Ref.
\cite{othera0}, in which an experimental analysis of the Crystal Barrel data
for the $a_{0}$ meson has been performed, see also Refs. \cite{bugg,reconc}.
However, we shall also point out that care is needed when extracting the
coupling constants of scalar-to-pseudoscalar from radiative $\phi$ decays
alone because of strong correlations of the parameters entering in the fit. In
particular, the difficulty is due to the the fact that, if the decay mechanism
(a) is dominant, an extraction of $c_{f_{0}KK}$ (and eventually $d_{f_{0}KK})$
from the line shapes of $\phi\rightarrow\gamma\pi^{0}\pi^{0}$ reaction is hard
because $c_{f_{0}KK}$ and $d_{f_{0}KK}$ appear only in the denominator of the
propagator of $f_{0}$ and are not directly proportional to decay amplitude(s).
(This is not the case for mechanism b, where the kaon-loop amplitudes are
directly proportional to $c_{f_{0}KK}$ and $d_{f_{0}KK}$). A similar
discussion holds in the $a_{0}$ case. For this reason we did not include the
coupling to kaons as free parameter of the fit but we repeated the analysis
for different values of the latter checking how the fit is affected.

As a comparison we also perform the fit using non-derivative couplings only
(that is we set $c_{f_{0}\pi\pi}=c_{f_{0}KK}=0$ and we leave $d_{f_{0}\pi\pi}$
and $d_{f_{0}KK}$ free) and we show that a bad description of data for the
$f_{0}$ meson is obtained. In the $a_{0}$ case ($c_{a_{0}\pi\eta}=c_{a_{0}%
KK}=0$ and $d_{a_{0}\pi\eta}$ and $d_{a_{0}KK}$ free), while a fit without
derivatives is still acceptable, the resulting parameters are not compatible
with the experimental results of Ref. \cite{othera0}. Thus, the need of
including derivative-like couplings is favoured by our study.

The paper is organized as follows: in the next Section we extend the formalism
of \cite{lupo} to the derivative interaction and we derive the theoretical
expressions for the lines shapes of $\phi$-decays. In the third Section we
present the fits to the experimental results and we study the correlation of
the parameters. In the fourth Section we drive our conclusions and outlook.

\section{$\phi$ decays within derivative interactions}

\subsection{1-Loop within derivative interactions}

\begin{figure}
[ptb]
\begin{center}
\includegraphics[
height=1.721in,
width=5.2053in
]%
{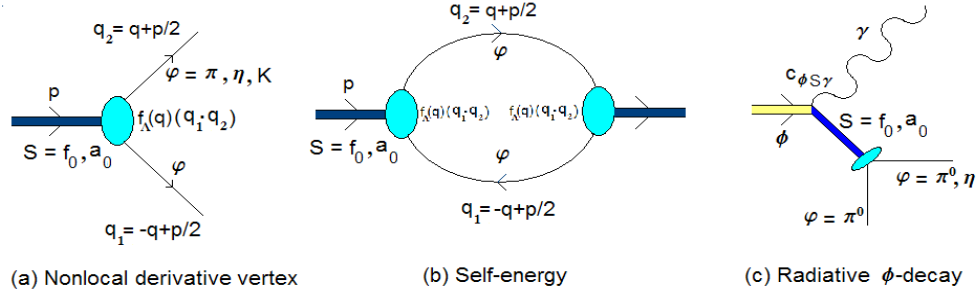}%
\caption{Relevant Feynman diagrams [colors online].}%
\label{feynmann}%
\end{center}
\end{figure}

We generalize the study of Ref. \cite{lupo} to which we refer for a careful
treatment of the definitions and relative discussions, by considering the
following nonlocal interaction Lagrangian of derivative type:%
\begin{equation}
\mathcal{L}=\frac{1}{2}(\partial_{\mu}S)^{2}-\frac{1}{2}M_{0}^{2}S^{2}%
+\frac{1}{2}(\partial_{\mu}\varphi)^{2}-\frac{1}{2}m^{2}\varphi^{2}%
+\mathcal{L}_{int}\text{ };\text{ }\mathcal{L}_{int}=gS(x)\int\mathrm{d}%
^{4}\mathrm{y}\partial_{\mu}\varphi(x+y/2)\partial^{\mu}\varphi(x-y/2)\Phi(y)
\end{equation}
with $\partial_{\mu}=\partial/\partial x^{\mu}$. The nonlocality describes the
finite dimensions of the scalar states \cite{lupo}: it takes into account
already at the Lagrangian level of a form factor in the expression of the
decay width. Previous studies (see, for instance, Refs.
\cite{amslerclose,tornqvist} and also the more microscopic approach of Ref.
\cite{godfrey}) show that a cut-off of the order of $1$ GeV emerges in the
context of phenomenological mesonic theories. By introducing the function
$f_{\Lambda}(q)$ as the Fourier transform of $\Phi(y),$ $f_{\Lambda}(q)=\int
d^{4}y\Phi(y)e^{-iyq},$ the tree-level Feynman amplitude of
Fig.~\ref{feynmann}.a reads $-ig\left(  q_{1}\cdot q_{2}\right)  f_{\Lambda
}\left(  (q_{1}-q_{2})/2\right)  $ where $q_{1}$ and $q_{2}$ are the momenta
of the two particles $\varphi$. Let $p=q_{1}+q_{2}$ be the momentum of the
particle $S$.%

The decay width is evaluated in the reference frame of $S,$ in which we have:%
\begin{equation}
p=(M_{0},\overrightarrow{0});\text{ }q_{1}=(\sqrt{\overrightarrow{q}^{2}%
+m^{2}},\overrightarrow{q});\text{ }q_{2}=(\sqrt{\overrightarrow{q}^{2}+m^{2}%
},-\overrightarrow{q});\text{ }q_{\varphi\varphi}(M_{0})=\left\vert
\overrightarrow{q}\right\vert =\sqrt{\frac{M_{0}^{2}}{4}-m^{2}}\text{ .}
\label{q(x)}%
\end{equation}
The tree-level decay reads explicitly:%
\begin{equation}
\Gamma_{S\varphi\varphi}^{\text{t-l}}(M_{0})=\frac{q_{\varphi\varphi}(M_{0}%
)}{8\pi M_{0}^{2}}\left[  (\sqrt{2}g)\left(  \frac{M_{0}^{2}-2m^{2}}%
{2}\right)  f_{\Lambda}(0,\overrightarrow{q})\right]  ^{2}\theta(M_{0}-2m)
\label{tl}%
\end{equation}
where the equality $q_{1}\cdot q_{2}=(M_{0}^{2}-2m^{2})/2$ has been used. The
tree-level propagator $\Delta_{S}(p)=\left[  p^{2}-M_{0}^{2}+i\varepsilon
\right]  ^{-1},$ valid in the limit $g\rightarrow0,$ is modified by the 1-loop
correction and takes the form $\Delta_{S}(x=\sqrt{p^{2}})=\left[  x^{2}%
-M_{0}^{2}+(\sqrt{2}g)^{2}\Sigma(x,m)+i\varepsilon\right]  ^{-1}$in which the
self-energy $\Sigma(x,m)$, see Fig. \ref{feynmann}.b, is given by:
\begin{equation}
\Sigma(x=\sqrt{p^{2}},m)=-i\int\frac{d^{4}q}{(2\pi)^{4}}\frac{\left[  \left(
q_{1}\cdot q_{2}\right)  f_{\Lambda}(q^{0},\overrightarrow{q})\right]  ^{2}%
}{\left[  q_{1}^{2}-m^{2}+i\varepsilon\right]  \left[  q_{2}^{2}%
-m^{2}+i\varepsilon\right]  } \label{int}%
\end{equation}
with $q_{1}=q+p/2$ and $q_{2}=-q+p/2.$ For future convenience we consider the
loop as a function of $x=\sqrt{p^{2}}$ and of the mass $m$. Note that the only
difference with respect to the non-derivative study of Ref. \cite{lupo} is the
extra-factor $\left(  q_{1}\cdot q_{2}\right)  ^{2}$ in the tree-level decay
rate and in the numerator of the self energy.

A possible choice of a three-dimensional cutoff, which makes the model finite,
corresponds to (\cite{amslerclose,tornqvist,godfrey}) $f_{\Lambda
}(q)=f_{\Lambda}(\overrightarrow{q}^{2})=\exp[-\overrightarrow{q}^{2}%
/\Lambda^{2}]$ with $q_{\varphi\varphi}(x)=\sqrt{\overrightarrow{q}^{2}}$ as
in Eq. (\ref{q(x)}). The precise form of the cutoff function does not
influence the results as long as convergence is achieved. Moreover, despite
the fact that the superficial degree of divergence of the integral in Eq.
(\ref{int}) in the limit of large $\Lambda$ is four, it turns out that the Eq.
(\ref{int}) is only linear divergent in $\Lambda.$ In our work we use
$\Lambda=1.5$ GeV, which is very close to the value used in Refs.
\cite{amslerclose,godfrey}. Variations of the latter between 1-2 GeV affects
only slightly the results.

\begin{figure}
[ptb]
\begin{center}
\includegraphics[
height=2.3186in,
width=4.5463in
]%
{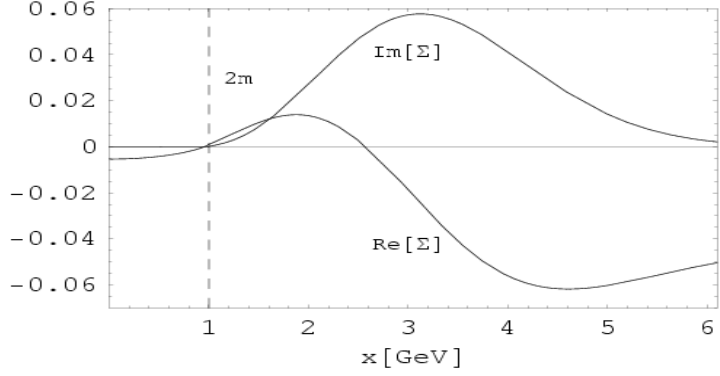}%
\caption{Real and imaginary parts of the self-energy $\Sigma(x)$ for the
illustrative values $m=0.5$ GeV and $\Lambda=1.5$ GeV.}%
\label{imre}%
\end{center}
\end{figure}

A general property for $\Sigma(p^{2})$ follows from the optical theorem:%
\begin{equation}
I_{S}(x)=\left(  \sqrt{2}g\right)  ^{2}\operatorname{Im}[\Sigma(x,m)]=x\Gamma
_{S\varphi\varphi}^{\text{t-l}}(x). \label{I(x)}%
\end{equation}
The imaginary part of the self-energy diagram is zero for $0<x<2m$ and nonzero
starting at threshold. The real part $R_{S}(x)=\left(  \sqrt{2}g\right)
^{2}\operatorname{Re}[\Sigma(x,m)]$ is nonzero below and above threshold. In
Fig. \ref{imre} the functions $\operatorname{Re}[\Sigma(x,m)]$ and
$\operatorname{Im}[\Sigma(x,m)]$ are plotted as an illustrative example.%

We define the (nominal) mass $M_{S}$ for the scalar field $S$ as the solution
of the equation $M_{S}^{2}-M_{0}^{2}+R_{S}(M_{S})=0.$ When the function
$R(M_{S})$ is positive, which is usually the physical case (Fig. 2), the
dressed mass $M_{S}$ is smaller than the bare mass $M_{0},$ showing that the
quantum fluctuations tend to lower it. We consider the case $M_{S}>2m,$ thus
no pole below threshold is found and we have a truly resonant state. The
spectral function $d_{S}(x)$ of the scalar field $S$ related to the imaginary
part of the propagator is:%
\begin{equation}
d_{S}(x)=\frac{2x}{\pi}\left\vert \lim_{\varepsilon\rightarrow0}%
\operatorname{Im}[\Delta_{S}(x)]\right\vert \overset{M_{S}>2m}{=}\frac{2x}%
{\pi}\frac{I_{S}(x)}{(x^{2}-M_{0}^{2}+R_{S}(x))^{2}+I_{S}(x)^{2}}.
\label{dsdef}%
\end{equation}
In the limit $g\rightarrow0$ we obtain the desired spectral function
$d_{S}(x)=\delta(x-M_{0}).$ The normalization of $d_{S}(x),$ i.e. the validity
of the K\"{a}llen-Lehman representation \cite{achasovprop}, holds for a large
range of $g$: $\int_{0}^{\infty}d_{S}(x)dx=1.$The decay rate for the process
$S\rightarrow\varphi\varphi$, which includes finite-width effects, can be
defined as $\Gamma_{S\varphi\varphi}=\int_{0}^{\infty}dxd_{S}(x)\Gamma
_{S\varphi\varphi}^{\text{t-l}}(x)$. This formula reduces to the tree-level
amplitude $\Gamma_{S\varphi\varphi}^{\text{t-l}}(M_{0})$ of Eq. (\ref{tl}) in
the limit of small $g$: $\Gamma_{S\varphi\varphi}^{\text{t-l}}(M_{0}%
)\simeq\Gamma_{S\varphi\varphi}$ for $g\rightarrow0$. We shall however not use
this formula in the present work: in fact, while mathematically correct, it
presents some practical mismatch in the case of derivative couplings because a
long-tail of the function $d_{S}(x)\Gamma_{S\varphi\varphi}^{\text{t-l}}(x)$
may arise at large $x$. We shall therefore not compare integrated decay widths
but decay amplitudes evaluated on shell, which are free form these ambiguities.

As a last step of this subsection we study the case of derivative interactions
of the scalar field $S$ with two different particles $\varphi_{1}$ and
$\varphi_{2}$ with masses $m_{1}$ and $m_{2}$: $\mathcal{L}_{int}%
=gS\partial_{\mu}\varphi_{1}\partial^{\mu}\varphi_{2}$ and its nonlocal
extension. The previous formulas change as follows:%
\begin{equation}
\Gamma_{S\varphi_{1}\varphi_{2}}^{\text{t-l}}(M_{0})=\frac{q_{\varphi
_{1}\varphi_{2}}(M_{0})}{8\pi M_{0}^{2}}\left[  g\left(  \frac{M_{0}^{2}%
-m_{1}^{2}-m_{2}^{2}}{2}\right)  f_{\Lambda}(0,\overrightarrow{q})\right]
^{2}\theta(M_{0}-m_{1}-m_{2})
\end{equation}
where $q_{\varphi_{1}\varphi_{2}}(M_{0})=\frac{1}{2M_{0}}\sqrt{M_{0}%
^{4}+(m_{1}^{2}-m_{2}^{2})^{2}-2(m_{1}^{2}+m_{2}^{2})M_{0}^{2}}.$ The
corresponding loop function is denoted as $\widetilde{\Sigma}(x,m_{1},m_{2})$
and reads%
\begin{equation}
\widetilde{\Sigma}(x,m_{1},m_{2})=-i\int\frac{d^{4}q}{(2\pi)^{4}}\frac{\left[
\left(  q_{1}\cdot q_{2}\right)  f_{\Lambda}(q^{0},\overrightarrow{q})\right]
^{2}}{\left[  q_{1}^{2}-m_{1}^{2}+i\varepsilon\right]  \left[  q_{2}^{2}%
-m_{2}^{2}+i\varepsilon\right]  }\text{ .}%
\end{equation}

\subsection{The $f_{0}(980)$ case}

\subsubsection{Derivative interaction with pions and kaons}

The field $f_{0}$, describing the resonance $f_{0}(980)$, interacts via
derivative couplings with pions and kaons. In the local limit one has:%
\begin{equation}
\mathcal{L}_{int,f_{0}}=c_{f_{0}\pi\pi}f_{0}\left(  \partial_{\mu
}\overrightarrow{\pi}\right)  ^{2}+c_{f_{0}KK}f_{0}\left(  \left(
\partial_{\mu}K^{+}\right)  \left(  \partial^{\mu}K^{-}\right)  +\left(
\partial_{\mu}K^{0}\right)  (\partial^{\mu}\overline{K}^{0})\right)
\end{equation}
where $c_{f_{0}\pi\pi}$ and $c_{f_{0}KK}$ are the coupling constants. The
nonlocal case, which is used in the following calculations, is obtained by
delocalizing the previous interaction Lagrangian:%

\begin{equation}
\mathcal{L}_{int,f_{0}}=gf_{0}(x)\int\mathrm{d}^{4}\mathrm{y}\partial_{\mu
}\overrightarrow{\pi}(x+y/2)\partial^{\mu}\overrightarrow{\pi}(x-y/2)\Phi
(y)+\text{ `kaon part'.}%
\end{equation}
The tree-level decay formulas, as function of the mass $x$, read:%
\begin{align}
\Gamma_{f_{0}\pi\pi}^{\text{t-l}}(x)  &  =\frac{q_{\pi\pi}(x)}{8\pi x^{2}%
}\left[  A_{f_{0}\pi\pi}(x)\right]  ^{2}\theta(x-2m_{\pi});\text{ }A_{f_{0}%
\pi\pi}(x)=\sqrt{6}c_{f_{0}\pi\pi}\left(  \frac{x^{2}-2m_{\pi}^{2}}{2}\right)
f_{\Lambda}(q_{\pi\pi}(x))\label{f0pipi}\\
\Gamma_{f_{0}KK}^{\text{t-l}}(x)  &  =\frac{q_{KK}(x)}{8\pi x^{2}}\left[
A_{f_{0}KK}(x)\right]  ^{2}\theta(x-2m_{K});\text{ }A_{f_{0}KK}(x)=\sqrt
{2}c_{f_{0}KK}\left(  \frac{x^{2}-2m_{K}^{2}}{2}\right)  f_{\Lambda}%
(q_{KK}(x))\text{ }, \label{f0kk}%
\end{align}
where for future use we introduced the amplitudes $A_{f_{0}\pi\pi}(x)$ and
$A_{f_{0}KK}(x)$. The field $f_{0}$ is dressed by pions and kaons (Fig.
\ref{feynmann}.b). Thus, the real and imaginary terms of the self-energy
include contributions from both loops:%

\begin{align}
I_{f_{0}}(x)  &  =\left(  \sqrt{6}c_{f_{0}\pi\pi}\right)  ^{2}%
\operatorname{Im}[\Sigma(x,m_{\pi})]+\left(  \sqrt{2}c_{f_{0}KK}\right)
^{2}\operatorname{Im}[\Sigma(x,m_{K})],\\
R_{f_{0}}(x)  &  =\left(  \sqrt{6}c_{f_{0}\pi\pi}\right)  ^{2}%
\operatorname{Re}[\Sigma(x,m_{\pi})]+\left(  \sqrt{2}c_{f_{0}KK}\right)
^{2}\operatorname{Re}[\Sigma(x,m_{K})].
\end{align}
The optical theorem holds for the single channels:%
\begin{equation}
\left(  \sqrt{6}c_{f_{0}\pi\pi}\right)  ^{2}\operatorname{Im}[\Sigma(x,m_{\pi
})]=x\Gamma_{f_{0}\pi\pi}^{\text{t-l}}(x),\text{ }\left(  \sqrt{2}c_{f_{0}%
KK}\right)  ^{2}\operatorname{Im}[\Sigma(x,m_{K})]=x\Gamma_{f_{0}%
KK}^{\text{t-l}}(x).
\end{equation}
The spectral function $d_{f_{0}}(x)$ is just as in Eq. (\ref{dsdef}) upon
setting $S=f_{0}$.

\subsubsection{$\phi\rightarrow\gamma\pi^{0}\pi^{0}$ decay}

Be $F^{\mu\nu}=\partial^{\mu}A^{\nu}-\partial^{\nu}A^{\mu}$ the
electromagnetic field strength and $V_{\mu\nu}=\partial_{\mu}\phi_{\nu
}-\partial_{\nu}\phi_{\mu}$ the field strength related to the vector field
$\phi_{\mu},$ which describes the resonance $\phi(1024)$ of the PDG
\cite{pdg}. The Lagrangian which describes the process $\phi\rightarrow
\gamma\pi^{0}\pi^{0}$ and the corresponding tree-level decay rate read
\begin{equation}
\mathcal{L}_{\phi\gamma f_{0}}=c_{\phi\gamma f_{0}}f_{0}F^{\mu\nu}V_{\mu\nu
}\rightarrow\Gamma_{\phi\gamma f_{0}}^{\text{t-l}}(x)=c_{\phi\gamma f_{0}}%
^{2}\frac{\left(  m_{\phi}^{2}-x^{2}\right)  ^{3}}{24\pi m_{\phi}^{3}}.
\end{equation}
When $f_{0}$ is on shell one sets $x=M_{f_{0}.}$ However, we are interested to
the subsequent decay of $f_{0}$ into $\pi^{0}\pi^{0},$ in which $f_{0}$ is a
virtual state as depicted in Fig. \ref{feynmann}.c. The partial decay rate,
defining the line shape of the $\phi\rightarrow\gamma\pi^{0}\pi^{0}$ decay,
reads%
\begin{equation}
\frac{d\Gamma_{\phi\gamma\pi^{0}\pi^{0}}(x)}{dx}=\Gamma_{\phi\gamma f_{0}%
}^{\text{t-l}}(x)\left[  \frac{2x}{\pi}\frac{x\Gamma_{f_{0}\pi^{0}\pi^{0}%
}^{\text{t-l}}(x)}{\left(  x^{2}-M_{0,f_{0}}^{2}+R_{f_{0}}(x)\right)
^{2}+I_{f_{0}}(x)^{2}}\right]  \label{f0ls}%
\end{equation}
where $\Gamma_{f_{0}\pi^{0}\pi^{0}}^{\text{t-l}}(x)=\frac{1}{3}\Gamma
_{f_{0}\pi\pi}^{\text{t-l}}(x).$ It can be also rewritten as%
\begin{equation}
d\Gamma_{\phi\gamma\pi^{0}\pi^{0}}(x)=\Gamma_{\phi\gamma f_{0}}^{\text{t-l}%
}(x)\left[  d_{f_{0}}(x)dx\right]  \left[  \frac{\Gamma_{f_{0}\pi^{0}\pi^{0}%
}^{\text{t-l}}(x)}{\Gamma_{f_{0}\pi\pi}^{\text{t-l}}(x)+\Gamma_{f_{0}%
KK}^{\text{t-l}}(x)}\right]  ,
\end{equation}
whose interpretation is straightforward: $\Gamma_{\phi\gamma f_{0}%
}^{\text{t-l}}(x)$ describes the decay rate for $\phi\rightarrow\gamma f_{0},$
$d_{f_{0}}(x)dx$ represents the probability that the particle $f_{0}$ has a
mass between $x$ and $x+dx$ and finally $\left[  \Gamma_{f_{0}\pi^{0}\pi^{0}%
}^{\text{t-l}}(x)/(\Gamma_{f_{0}\pi\pi}^{\text{t-l}}(x)+\Gamma_{f_{0}%
KK}^{\text{t-l}}(x))\right]  $ describes the branching ratio of $f_{0}$
decaying into a $\pi^{0}\pi^{0}$ pair. The quantity $d\Gamma_{\phi\gamma
\pi^{0}\pi^{0}}(x)/dx$ can be directly compared with experiments as we do in
the next section \cite{fn1}.

\subsection{The $a_{0}(980)$ case}

\subsubsection{Derivative interaction with $\pi^{0}\eta$ and kaons}

The discussion concerning the resonance $a_{0}(980)$ follows the same line of
the previous subsection. The interaction Lagrangian for the neutral field
$a_{0}^{0}\equiv a_{0}^{0}(980)$ reads in the local limit
\begin{equation}
\mathcal{L}_{int,a_{0}^{0}}=c_{a_{0}\pi\eta}a_{0}^{0}\left(  \partial_{\mu}%
\pi^{0}\right)  \left(  \partial_{\mu}\eta\right)  +c_{a_{0}KK}a_{0}%
^{0}\left(  \left(  \partial_{\mu}K^{+}\right)  \left(  \partial^{\mu}%
K^{-}\right)  +\left(  \partial_{\mu}K^{0}\right)  (\partial^{\mu}\overline
{K}^{0})\right)  .
\end{equation}
where the coupling constants $c_{a_{0}\pi\eta}$ and $c_{a_{0}KK}$ have been
introduced. The tree-level decay rates in the nonlocal case are%
\begin{align}
\Gamma_{a_{0}\pi\eta}^{\text{t-l}}(x)  &  =\frac{q_{\pi\eta}(x)}{8\pi x^{2}%
}\left[  A_{a_{0}\pi\eta}(x)\right]  ^{2}\theta(x-m_{\pi}-m_{\eta});\text{
}A_{a_{0}\pi\eta}(x)=c_{a_{0}\pi\eta}\left(  \frac{x^{2}-m_{\pi}^{2}-m_{K}%
^{2}}{2}\right)  f_{\Lambda}(q_{\pi\eta}(x))\label{a0pieta}\\
\Gamma_{a_{0}KK}^{\text{t-l}}(x)  &  =\frac{q_{KK}(x)}{8\pi x^{2}}\left[
A_{a_{0}KK}(x)\right]  ^{2}\theta(x-2m_{K});\text{ }A_{a_{0}KK}(x)=\sqrt
{2}c_{a_{0}KK}\left(  \frac{x^{2}-2m_{K}^{2}}{2}\right)  f_{\Lambda}%
(q_{KK}(x)) \label{a0kk}%
\end{align}
where the amplitudes $A_{a_{0}\pi\eta}(x)$ and $A_{a_{0}KK}(x)$ have been
introduced. The real and imaginary parts of the loop include contributions
from $\pi\eta$ and $KK$ loops:%

\begin{align}
I_{a_{0}}(x)  &  =\left(  c_{a_{0}\pi\eta}\right)  ^{2}\operatorname{Im}%
[\widetilde{\Sigma}(x,m_{\pi},m_{\eta})]+\left(  \sqrt{2}c_{a_{0}KK}\right)
^{2}\operatorname{Im}[\Sigma(x,m_{K})];\\
R_{a_{0}}(x)  &  =\left(  c_{a_{0}\pi\eta}\right)  ^{2}\operatorname{Re}%
[\widetilde{\Sigma}(x,m_{\pi},m_{\eta})]+\left(  \sqrt{2}c_{a_{0}KK}\right)
^{2}\operatorname{Re}[\Sigma(x,m_{K})].
\end{align}
The spectral function $d_{a_{0}}(x)$ reads as in Eq. (\ref{dsdef}) upon
setting $S=a_{0}.$

\subsubsection{$\phi\rightarrow\gamma\pi^{0}\eta$ decay}

The interaction Lagrangian describing the process $\phi\rightarrow\gamma
\pi^{0}\eta$ and the corresponding tree-level decay rate are given by:
\begin{equation}
\mathcal{L}_{\phi\gamma a_{0}}=c_{\phi\gamma a_{0}}a_{0}^{0}F^{\mu\nu}%
V_{\mu\nu}\text{ }\rightarrow\Gamma_{\phi\gamma a_{0}}^{\text{t-l}}%
(x)=c_{\phi\gamma a_{0}}^{2}\frac{\left(  m_{\phi}^{2}-x^{2}\right)  ^{3}%
}{24\pi m_{\phi}^{3}}%
\end{equation}
where $x=M_{a_{0}}$ for an on-shell decay. However, the $a_{0}^{0}$ meson
decays subsequently into $\pi^{0}\eta$ as depicted in Fig. \ref{feynmann}.c.
As a result the line shape of the reaction $\phi\rightarrow\gamma\pi^{0}\eta$
reads:
\begin{equation}
\frac{d\Gamma_{\phi\gamma\pi^{0}\eta}(x)}{dx}=\Gamma_{\phi\gamma a_{0}%
}^{\text{t-l}}(x)\left[  \frac{2x}{\pi}\frac{x\Gamma_{a_{0}\pi\eta
}^{\text{t-l}}(x)}{\left(  x^{2}-M_{0,a_{0}}^{2}+R_{a_{0}}(x)\right)
^{2}+I_{a_{0}}(x)^{2}}\right]  \text{ }.
\end{equation}

\section{Fit of the line shapes}

\subsection{The $\phi\rightarrow\gamma\pi^{0}\pi^{0}$ fit}

\begin{figure}
[ptb]
\begin{center}
\includegraphics[
height=2.9456in,
width=4.2454in
]%
{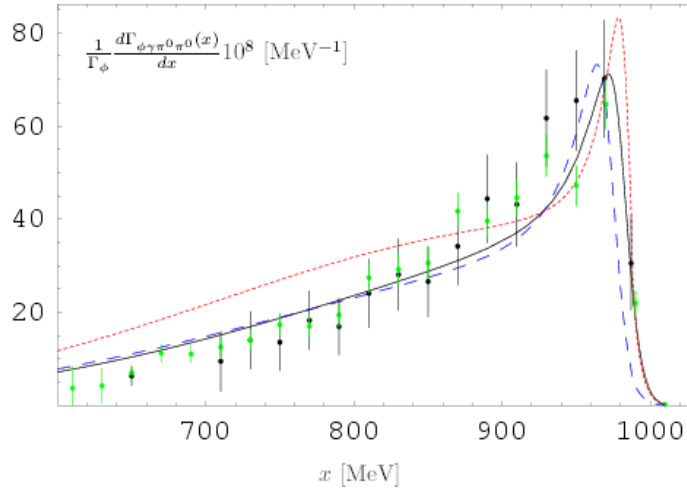}%
\caption{Branching ratio $\frac{1}{\Gamma_{\phi}}\frac{d\Gamma_{\phi\gamma
\pi^{0}\pi^{0}}(x)}{dx}10^{8}$ [MeV$^{-1}$] as function of the invariant mass
$x$ [colors online]. $\Gamma_{\phi}=4.26$ MeV is the full width of the $\phi$
meson. We consider data sets from the SND and KLOE collaborations
\cite{sndf0,kloef0} corresponding respectively to the black and grey (green
online) dots. The continuous line is the result of the fit by setting
$c_{f_{0}KK}=0,$ the dashed line corresponds to the case $c_{f_{0}KK}=12$
GeV$^{-1}$. Both cases are in Table I. The dotted line corresponds also to
$c_{f_{0}KK}=12$ GeV$^{-1}$ but only data points above 0.8 GeV are used in the
fit, see Table II.}%
\label{fitf0}%
\end{center}
\end{figure}

Five parameters $\Lambda,M_{f_{0}},c_{\phi\gamma f_{0}},c_{f_{0}\pi\pi
},c_{f_{0}KK}$ determine the $\phi\rightarrow\gamma\pi^{0}\pi^{0}$ line shape
via a virtual $f_{0}(980)$ meson \cite{fn2}. We fix the cutoff $\Lambda=1.5$
GeV; a mild dependence of the results is seen by varying $\Lambda$. Being the
coupling to kaons of particular theoretical interest we perform the fit for
different values of $c_{{f_{0}}KK}$. We start setting $c_{{f_{0}}KK}=0$ ; the
fit is done to the experimental data of SND and KLOE collaborations
\cite{sndf0,kloef0} corresponding to the black and grey dots in
Fig.~\ref{fitf0} retaining the data points above $0.6$ GeV. The solid line is
the corresponding theoretical curve for $c_{{f_{0}}KK}=0$, whose fit
parameters are listed in the first entry of Table \ref{tab1}. The peak of the
$f_{0}$ line shape is a feature which is easily reproduced by derivative interactions.

Let us now investigate how the fit changes by increasing $c_{{f_{0}}KK}$ from
$0$ to a maximum value of $14$ GeV$^{-1}$. As we can see in Table \ref{tab1},
the fitted mass varies slowly from $M_{f_{0}}\thicksim981$ MeV to $M_{f_{0}%
}\thicksim971$ MeV, $c_{f_{0}\pi\pi}$ increases from 1.28 to 2.90 GeV$^{-1}%
$and the $\frac{\chi^{2}}{d.o.f.}$ also increases from 1.7 to 4. We also
present the on-shell amplitudes $A_{f_{0}\pi\pi}=A_{f_{0}\pi\pi}(M_{f_{0}})$
and $A_{f_{0}KK}=A_{f_{0}KK}(M_{f_{0}})$ defined in Eq. (\ref{f0pipi}%
)-(\ref{f0kk}), which in the following will be compared with the results of
\cite{otherf0}. The value of $\chi^{2}/d.o.f$ increases by increasing
$c_{{f_{0}}KK}$. However, as already mentioned in the Introduction, the
coupling constant $c_{f_{0}KK}$ appears only in the denominator of the
propagator of the $f_{0}$ meson (see Eq. (\ref{f0ls})) and therefore its
determination is difficult. Nevertheless, we notice that rather acceptable
fits are found when varying $c_{{f_{0}}KK}$ in such large range. We thus not
intend to state that $c_{{f_{0}}KK}$ is small (what would be in conflict with
data from $\pi\pi$ scattering) but only that its determination from the line
shape is problematic when mechanism a.1 (whose amplitude is proportional to
$c_{f_{0}\pi\pi})$ is regarded as dominant.%

\begin{table}[ptb]
\caption{Fit of the $\phi\rightarrow\gamma\pi^{0}\pi^{0}$ line shape data from
$x>600$ MeV.}%
\label{tab1}
\begin{tabular}
[c]{|c|c|c|c|c|c|c|}\hline
$%
\begin{array}
[c]{c}%
c_{f_{0}KK}\\
(GeV^{-1})
\end{array}
$ & $%
\begin{array}
[c]{c}%
M_{f_{0}}\\
(MeV)
\end{array}
$ & $%
\begin{array}
[c]{c}%
c_{f_{0}\pi\pi}\\
(GeV^{-1})
\end{array}
$ & $%
\begin{array}
[c]{c}%
c_{\phi\gamma f_{0}}\\
(GeV^{-1})
\end{array}
$ & $\frac{\chi^{2}}{d.o.f.}$ & $%
\begin{array}
[c]{c}%
A_{f_{0}KK}\\
(GeV)
\end{array}
$ & $%
\begin{array}
[c]{c}%
A_{f_{0}\pi\pi}\\
(GeV)
\end{array}
$\\\hline
$0$ & $981.8\pm2.8$ & $1.28\pm0.04$ & $0.249\pm0.008$ & $1.7$ & $0$ &
$1.31\pm0.042$\\\hline
$2$ & $981.7\pm2.8$ & $1.30\pm0.04$ & $0.260\pm0.008$ & $1.9$ & $0.675\pm0.01$
& $1.34\pm0.045$\\\hline
$4\ $ & $981.7\pm2.8$ & $1.38\pm0.05$ & $0.291\pm0.010$ & $2.3$ &
$1.35\pm0.01$ & $1.42\pm0.06$\\\hline
$8$ & $974.4\pm2.6$ & $2.00\pm0.06$ & $0.312\pm0.04$ & $3.4$ & $2.62\pm0.03$ &
$2.02\pm0.06$\\\hline
$12\ $ & $971.8\pm2.3$ & $2.59\pm0.06$ & $0.383\pm0.011$ & $3.9$ &
$3.89\pm0.04$ & $2.61\pm0.06$\\\hline
$14$ & $971.2\pm2.3$ & $2.90\pm0.07$ & $0.425\pm0.11$ & $4.0$ & $4.53\pm0.04$
& $2.92\pm0.07$\\\hline
\end{tabular}
\end{table}

Notice also that at low $x$ the resonance $f_{0}(600)$ contributes to the
total branching ratio as found in Refs. \cite{gubin,bugg}. For this reason we
repeated the analysis by retaining only the data points above 800 MeV where
$f_{0}(600)$ is less relevant. The corresponding results are presented in
Table \ref{tab2}, where significantly smaller $\chi^{2}$ are obtained. Also
for large values of $c_{{f_{0}}KK}$ the results are satisfying. As we increase
$c_{{f_{0}}KK}$ from $0$ to $14$ GeV$^{-1}$, the fitted mass is almost
unchanged $M_{f_{0}}\thicksim984$ MeV, $c_{f_{0}\pi\pi}$ increases from 1.35
to 2.35 GeV$^{-1}$and the $\frac{\chi^{2}}{d.o.f.}$ increases from 0.9 to 2.
In Fig. \ref{fitf0} the dotted line corresponds to the case $c_{{f_{0}}KK}=12$
GeV$^{-1}$ of Table \ref{tab2}: while the data above 0.8 GeV are well
described, an overestimation of data point between 0.6 and 0.8 GeV is clearly
visible. Thus, in the present analysis a destructive interference with
$\sigma$ meson should occur: this fact can represent a constraint on models of
light scalar mesons. Interestingly, a destructive interference of $f_{0}(600)$
and $f_{0}(980)$ is also the outcome of Ref.~\cite{kloef0}, where the
$f_{0}(980)$ channel overestimates the data below $\sim700$ MeV. Nevertheless
a more refined analysis including the interference with the $f_{0}(600)$ meson
and using the new data of KLOE \cite{kloenew} -for which at present no tables
with branching ratios have been presented- will be compulsory and represents
an outlook of the present work. At this stage a preliminary comparison with
the new data of KLOE can be done only by considering the integrated branching
ratio given in Ref.~\cite{kloenew}: $BR(\phi\rightarrow\pi^{0}\pi^{0}%
\gamma)=1.07\times10^{-4}$. Our value runs from $1.06-1.10\times10^{-4}$,
depending on the choice for $c_{f_{0}KK}$, and therefore is consistent with
the latest experimental results.

\begin{table}[ptb]
\caption{Fit of the $\phi\rightarrow\gamma\pi^{0}\pi^{0}$ line shape data from
$x>800$ MeV.}%
\label{tab2}%
\begin{tabular}
[c]{|c|c|c|c|c|c|c|}\hline
$%
\begin{array}
[c]{c}%
c_{f_{0}KK}\\
(GeV^{-1})
\end{array}
$ & $%
\begin{array}
[c]{c}%
M_{f_{0}}\\
(MeV)
\end{array}
$ & $%
\begin{array}
[c]{c}%
c_{f_{0}\pi\pi}\\
(GeV^{-1})
\end{array}
$ & $%
\begin{array}
[c]{c}%
c_{\phi\gamma f_{0}}\\
(GeV^{-1})
\end{array}
$ & $\frac{\chi^{2}}{d.o.f.}$ & $%
\begin{array}
[c]{c}%
A_{f_{0}KK}\\
(GeV)
\end{array}
$ & $%
\begin{array}
[c]{c}%
A_{f_{0}\pi\pi}\\
(GeV)
\end{array}
$\\\hline
$0$ & $984.2\pm3.2$ & $1.35\pm0.04$ & $0.263\pm0.007$ & $0.9$ & $0$ &
$1.39\pm0.04$\\\hline
$2$ & $984.1\pm3.0$ & $1.38\pm0.04$ & $0.275\pm0.007$ & $0.9$ & $0.68\pm0.01$
& $1.42\pm0.04$\\\hline
$4\ $ & $984.0\pm2.8$ & $1.48\pm0.05$ & $0.309\pm0.009$ & $1.0$ &
$1.36\pm0.01$ & $1.52\pm0.05$\\\hline
$8$ & $983.6\pm2.4$ & $1.81\pm0.07$ & $0.419\pm0.013$ & $1.5$ & $2.72\pm0.03$
& $1.86\pm0.07$\\\hline
$12\ $ & $983.4\pm2.4$ & $2.25\pm0.09$ & $0.558\pm0.019$ & $1.9$ &
$4.07\pm0.04$ & $2.32\pm0.09$\\\hline
$14$ & $983.3\pm2.3$ & $2.50\pm0.11$ & $0.632\pm0.02$ & $2.0$ & $4.75\pm0.04$
& $2.57\pm0.11$\\\hline
\end{tabular}
\par
\bigskip\end{table}We now compare and discuss our amplitudes extracted from
the KLOE and SND data with the amplitudes extracted from the experimental
analyses of \cite{otherf0} for the $f_{0}$ meson via $j/\psi$ decay at BES.
Let us stress that a comparison by using the decay widths would be less
reliable, since they depend on the adopted way to evaluate them. The
amplitudes extracted in Refs. \cite{otherf0,bugg} \ are :%
\begin{equation}
A_{f_{0}\pi\pi}=2.88\pm0.22\text{ GeV},\text{ }A_{f_{0}KK}=5.91\pm0.77\text{
GeV.} \label{amplf0}%
\end{equation}
Our last entries in Tables \ref{tab1} and \ref{tab2} are in qualitative
agreement with Eq. (\ref{amplf0}). A more quantitative check is possible: we
deduce from Eq. (\ref{amplf0}) the couplings $c_{f_{0}\pi\pi}$ and
$c_{f_{0}KK}$ (at a fixed value of the mass of the meson) and use them in the
fit of the KLOE and SND data leaving the coupling $c_{\phi\gamma f_{0}}$ as
the only free parameter. The fit turns out to be acceptable with $\frac
{\chi^{2}}{d.o.f.}\thicksim2.4$, the extracted values of the couplings are
$c_{f_{0}\pi\pi}=2.79$ GeV$^{-1}$ and $c_{f_{0}KK}=17.37$ GeV$^{-1}$ for
$M_{f_{0}}=984$ MeV. The fact that the line shape of $\phi\rightarrow\gamma
\pi^{0}\pi^{0}$ can be well described by using Eq. (\ref{amplf0}) as an input
shows that the BES results of Ref. \cite{otherf0} are compatible with our analysis.

\subsection{The $\phi\rightarrow\gamma\pi^{0}\eta$ fit}

\begin{figure}
[ptb]
\begin{center}
\includegraphics[
height=2.9447in,
width=4.5809in
]%
{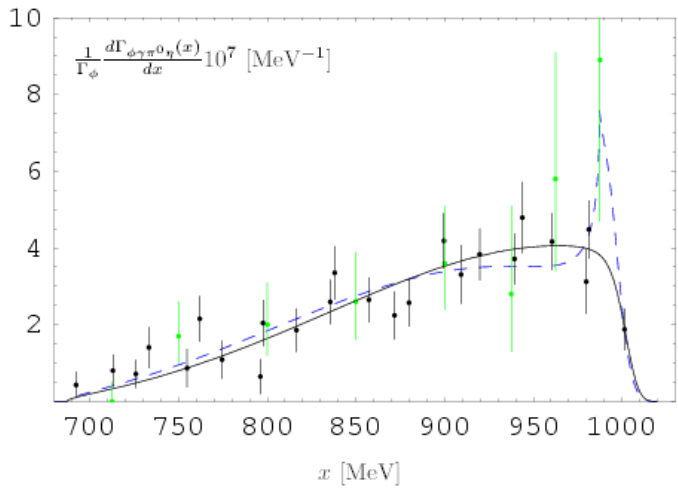}%
\caption{Branching ratio $\frac{1}{\Gamma_{\phi}}\frac{d\Gamma_{\phi\gamma
\pi^{0}\eta}(x)}{dx}10^{7}$ [MeV$^{-1}$] as function of the invariant mass $x$
[colors online]. $\Gamma_{\phi}=4.26$ MeV is the full width of $\phi$ meson.
Grey points (green online) from \cite{snda0} and black ones from
\cite{kloea0}. The solid line corresponds to $c_{a_{0}KK}=0$ while the dashed
one -with the pronounced peak at threshold- to $c_{a_{0}KK}=12$ GeV$^{-1}$.}%
\label{fita0}%
\end{center}
\end{figure}

We proceed as in the $f_{0}$ case by first setting $c_{a_{0}KK}=0$: the fitted
curve is shown by the solid line in Fig.~\ref{fita0}. We then increase
$c_{a_{0}KK}$ up to $14$ GeV$^{-1}$: the fit turns out to be satisfying in the
whole range, see Table \ref{tab3} where the mass, the couplings and the
on-shell amplitudes $A_{a_{0}\pi\eta}=A_{a_{0}\pi\eta}(M_{a_{0}})$,
$A_{a_{0}KK}=A_{a_{0}KK}(M_{a_{0}})$ defined in Eq. (\ref{a0pieta}%
)-(\ref{a0kk}), are reported. In Fig.~\ref{fita0} we also show the fitted
curve in the case $c_{a_{0}KK}=12$ GeV$^{-1}$ (dashed line). The narrow peak
due to threshold effects is remarkable, but the data are not precise enough to
determine its existence.

A comment about the behavior of the $\frac{\chi^{2}}{d.o.f.}$ as $c_{a_{0}KK}$
varies is in order: the fact that the fit is quite good for all the values of
$c_{a_{0}KK}$ is a signal of a strong correlation between the parameters. As
in the $f_{0}$ case $c_{a_{0}KK}$ only appears in the denominator of the
propagator, thus the dependence of the line shape on this parameter is weak
and its determination hard. While the value of the mass is almost constant,
$c_{a_{0}\pi\eta}$ and $c_{\phi\gamma a_{0}}$ change significantly. This means
that, while the fitted curve is in agreement with the data, the determination
of the parameters is not reliable. This is due to the large error bars of the
data, which in turn produce also large errors associated with each of the
parameters of the fit. Again a new analysis with the new data of KLOE would be
extremely important. As before a comparison with the new results of KLOE on
the integrated branching ratio is useful. In Ref.~\cite{kloenew} it is found
that $BR(\phi\rightarrow\eta\pi^{0}\gamma)=6.92-7.19\times10^{-5}$
(respectively for the chains $\eta\rightarrow\gamma\gamma$ and $\eta
\rightarrow\pi^{+}\pi^{-}\pi^{0}$). Our value runs from $7.37-7.73\times
10^{-5}$, depending on the choice for $c_{a_{0}KK}$, and therefore is in
reasonable agreement with the latest experimental results.

We now compare our results with the amplitudes extracted from the experimental
analysis of the Crystal Barrel data of Ref. \cite{othera0}:%

\begin{equation}
A_{a_{0}\pi\eta}=3.33\pm0.15\text{ GeV},\text{ }A_{a_{0}KK}=3.59\pm0.44\text{
GeV}, \label{ampla0}%
\end{equation}
which are not far from the last three entries of Table \ref{tab3}. As done
previously, a more quantitative test consists in deducing the couplings
$c_{a_{0}\pi\eta}$ and $c_{a_{0}KK}$ from\ Eq. (\ref{ampla0}) and, at a fixed
value of the mass, perform the fit to SND and KLOE data with only
$c_{\phi\gamma a_{0}}$ as free parameter. The fit turns out to reproduce
correctly the data with $\frac{\chi^{2}}{d.o.f.}\thicksim1.2$, the extracted
values of the couplings are $c_{a_{0}\pi\eta}=10.17$ GeV$^{-1}$ and
$c_{a_{0}KK}=9.79$ GeV$^{-1}$ for $M_{a_{0}}=1004$ MeV. Also in this case we
obtain that our analysis is compatible with experimental results of a
different source.%

\subsection{Comparison with the fit without derivatives}

As we have seen a structureless approach using derivative couplings describes
well the KLOE and SND data of the $\phi$ radiative decays. This is manly due
to the fact that within the derivative approach one has an extra dependence of
the form $(x^{2}-m_{1}^{2}-m_{2}^{2})^{2}$ in the numerator of the line shape,
which emphasizes the peak of the distribution at large values of the invariant
mass. Since this term is not present in the non-derivative case the
description of the peaked line shapes more difficult. As a comparison we also
discuss in the following the results obtained when non-derivative interactions
are employed. One could still obtain the peaks with non-derivative couplings
by using very small decay amplitudes, which however are not in agreement with
the large amplitudes extracted from Refs. \cite{otherf0,othera0}, reported in
Eqs. (\ref{amplf0}) and (\ref{ampla0}).

More specifically, the inconsistency of the use of non-derivative couplings is
evident in the case of the $f_{0}$ meson: when setting $c_{f_{0}KK}%
=c_{f_{0}\pi\pi}=0$ and allowing for nonzero $d_{f_{0}KK}$ and $d_{f_{0}\pi
\pi}$ (see discussion in the Introduction) high $\frac{\chi^{2}}%
{d.o.f.}\gtrsim12$ are obtained for fits above 800 MeV.

When setting in the $a_{0}$ channel $c_{a_{0}KK}=c_{a_{0}\pi\eta}=0$ and
allowing for nonzero $d_{a_{0}KK}$ and $d_{a_{0}\pi\eta}$ fits with still
acceptable values of $\frac{\chi^{2}}{d.o.f.}\sim2.4$ are obtained, but the
resulting amplitude $A_{a_{0}\pi\eta}\sim1.37$ is smaller than all the cases
corresponding to the derivative coupling (see Table \ref{tab3}) and therefore
the comparison with the amplitudes of Eq. (\ref{ampla0}) is problematic.
Reversing the argument, we may deduce the values of $d_{a_{0}\pi\eta}$ and
$d_{a_{0}KK}$ from Eq. (\ref{ampla0}) (again fixing the mass of the meson) and
use them to fit the KLOE and SND data with one free coupling $c_{\phi\gamma
a_{0}}$. The obtained line shape describes very badly the data with
$\frac{\chi^{2}}{d.o.f.}\sim10$, much larger than all the cases corresponding
to derivative coupling. Thus, incompatibility with Eq. (\ref{ampla0}) is manifest.

All these results indicate that the use of non-derivative coupling is
disfavored by present experimental informations.

\begin{center}
\begin{table}[ptb]
\caption{Fit of the $\phi\rightarrow\gamma\pi^{0}\eta$ line shape.}%
\label{tab3}%
\begin{tabular}
[c]{|c|c|c|c|c|c|c|}\hline
$%
\begin{array}
[c]{c}%
c_{a_{0}KK}\\
(GeV^{-1})
\end{array}
$ & $%
\begin{array}
[c]{c}%
M_{a_{0}}\\
(MeV)
\end{array}
$ & $%
\begin{array}
[c]{c}%
c_{a_{0}\pi\eta}\\
(GeV^{-1})
\end{array}
$ & $%
\begin{array}
[c]{c}%
c_{\phi\gamma a_{0}}\\
(GeV^{-1})
\end{array}
$ & $\frac{\chi^{2}}{d.o.f.}$ & $%
\begin{array}
[c]{c}%
A_{a_{0}KK}\\
(GeV)
\end{array}
$ & $%
\begin{array}
[c]{c}%
A_{a_{0}\pi\eta}\\
(GeV)
\end{array}
$\\\hline
$0$ & $1005\pm18$ & $4.68\pm0.90$ & $0.263\pm0.046$ & $0.88$ & $0$ &
$1.54\pm0.3$\\\hline
$2$ & $1005\pm18$ & $4.71\pm0.94$ & $0.278\pm0.050$ & $0.88$ & $0.74\pm0.05$ &
$1.55\pm0.32$\\\hline
$4$ & $1004\pm18$ & $4.80\pm1.07$ & $0.321\pm0.06$ & $0.88$ & $1.47\pm0.10$ &
$1.57\pm0.36$\\\hline
$8$ & $1003\pm18$ & $5.10\pm1.62$ & $0.486\pm0.13$ & $0.88$ & $2.92\pm0.20$ &
$1.67\pm0.53$\\\hline
$12$ & $1002\pm19$ & $5.52\pm2.49$ & $0.731\pm0.28$ & $0.92$ & $4.36\pm0.31$ &
$1.80\pm0.82$\\\hline
$14$ & $1001\pm20$ & $5.77\pm3.01$ & $0.873\pm0.400$ & $0.92$ & $5.09\pm0.37$
& $1.88\pm1.00$\\\hline
\end{tabular}
\end{table}

\end{center}

\section{Conclusions}

In this work the role of derivative interactions of the mesons $f_{0}(980)$
and $a_{0}(980)$ with the pseudoscalar Goldstone bosons, which are a basic
consequence of spontaneous chiral symmetry breaking as shown explicitly in
chiral perturbation theory, has been investigated within a nonlocal approach
in relation to the radiative decays $\phi\rightarrow\gamma\pi^{0}\pi^{0}$ and
$\phi\rightarrow\gamma\pi^{0}\eta.$ After developing the method at one-loop
level it has been shown that a satisfactory description of the line shapes of
the reactions $\phi\rightarrow\gamma\pi^{0}\pi^{0}$ and $\phi\rightarrow
\gamma\pi^{0}\eta$ is obtained: the property of derivatives interactions seems
tailor-made to describe the well-marked peaks measured experimentally.

More specifically, in the $\phi\rightarrow\gamma\pi^{0}\pi^{0}$ case we fitted
our theoretical curves, which make use of derivative interactions only and
involve a virtual $f_{0}$ meson, to data above 600 and 800 MeV (Fig.
\ref{fitf0} and Tables \ref{tab1} and \ref{tab2}): in the latter case the
influence of the $f_{0}(600)$ resonance is less relevant. It has been also
stressed that the determination of the coupling to kaons from the line shape
only is hard because the latter enters only in the propagator of the scalar
meson. Interestingly, fits involving a large $f_{0}\rightarrow\overline{K}K$
coupling, as found in \cite{otherf0}, are acceptable ($\chi^{2}/d.o.f.$
$\lesssim2$) when data-points above 800 MeV are considered. Notice that in
this case data below 800 MeV are overestimated as the dotted line in Fig.
\ref{fitf0} shows. A destructive interference in the $f_{0}(600)$ channel
should be invoked in order to get agreement with data, see also Ref.
\cite{kloef0}. While leaving a more detailed study as an outlook, this fact
can represent a significant constraint on models based on a particular
interpretation of the scalar mesons, such as the tetraquark assignment. We
also verified that the usage of non-derivative couplings generates a bad
agreement with the experimental results ($\chi^{2}/d.o.f.$ $>10$): the line
shape of the $f_{0}$ meson cannot be reproduced for any choice of the parameters.

In the $\phi\rightarrow\gamma\pi^{0}\eta$ case, occurring via a virtual
$a_{0}$ meson, a good agreement is found (Fig. \ref{fita0} and Table
\ref{tab3}). However, even sizable variations of the parameters do not make
the agreement worse: for this reason a determination of the parameters is
hard. As above, the determination of the coupling to kaons is problematic.
However, when using the amplitudes extracted from Ref. \cite{othera0} to
deduce the couplings with pseudoscalar mesons a good description of the line
shape is obtained: this fact shows compatibility of the study of Crystal
Barrel data in Ref. \cite{othera0} with KLOE-SND data when derivative
interactions are used (notice also the pronounced peak close to threshold of
the dashed line in Fig. \ref{fita0}, whose appearance is due to a large
coupling to kaons). On the contrary, the same procedure in the non-derivative
case shows that the line shape of $\phi\rightarrow\gamma\pi^{0}\eta$ is badly reproduced.

Summarizing, the results of the present study point out that derivative-type
interactions (denoted as mechanism a.1 in the Introduction) with pseudoscalar
mesons can play an important role for the study of the scalar states
$a_{0}(980)$ and $f_{0}(980).$ The next step of the present work is the
inclusion of non-derivative interactions (explicit chiral symmetry breaking
terms, mechanism a.2) besides the derivative ones. The next-to-next (more
ambitious) step is the inclusion at the same time and in the same theoretical
framework, together with the mechanisms a.1 (derivative couplings) and a.2
(non-derivative couplings), also of the correspondent kaon-loop driven
contributions (mechanisms b.1 and b.2 respectively). To achieve these goals,
which can also set up which mechanism (if any) is dominant in radiative $\phi$
decays, further work is required. The aim is a definitive determination of the
relevant amplitudes, which represent a necessary tool to test models and thus
to understand the nature of the light scalar mesons. To this end the use of
other experimental informations from $j/\psi$ decay and Crystal Barrel data is
necessary. Also, a comparison with the newest KLOE data \cite{kloenew} and the
study of radiative decays of scalar resonances \cite{rad} represents an
interesting development.

\bigskip

\textbf{Acknowledgements} We thank D. Bugg and C. Bini for useful discussions.
G.P. acknowledges financial support from INFN.


\begin{thebibliography}{99}
\bibitem {fn1}There is another process with the same initial and final states
which can contribute to the line shape: $\phi\rightarrow\rho^{0}\pi
^{0}\rightarrow\pi^{0}\pi^{0}\gamma$. This process has been included to fit
the data of KLOE and SND in Refs.~\cite{sndf0,kloef0}, however it was found
that its contribution is small. Thus, we prefer, for the sake of simplicity,
to neglect this term in this preliminary investigation on the derivative interactions.
\bibitem {fn2}Notice that within our formalism we can consider $c_{f_{0}\pi
\pi}$ as a fit-parameter at variance from Ref.~\cite{black} where only the
product $c_{\phi\gamma f_{0}}c_{f_{0}\pi\pi}$ appears in the theoretical
curve. Therefore only $c_{\phi\gamma f_{0}}$ is used as a fit-parameter while
$c_{f_{0}\pi\pi}$ is fixed.
\end{thebibliography}
\end{document}